\documentclass[12pt,thmsa]{article}
\usepackage{amsfonts}
\usepackage{graphicx}
\usepackage{epsfig}
\usepackage{graphics}

\makeatletter
\newcommand{\row}[1]{\mathord{\buildrel{\lower3pt\hbox{$\scriptscriptstyle\rightarrow$}}\over #1}}

\newcommand{\dyadic}[1]{\mathord{\dyadic@rrow{#1}}}
\newcommand{\dyadic@rrow}[1]{
\begin{picture}(12,12)(-1,0)
\put(-3,12){\makebox(0,0)[t]{$\scriptscriptstyle\downarrow$}}
\put(-3,13){\makebox(0,0)[l]{$\scriptscriptstyle\longrightarrow$}}
\put(5,0){\makebox(0,0)[b]{$#1$}}
\end{picture}
}
\newcommand{\bra}[1]{\bigl\langle #1 \bigr|}
\newcommand{\ket}[1]{\bigl| #1 \bigr\rangle}

\input{tcilatex}

\begin{document}

\date{\today}

\begin{center}
{ Entanglement and quantum teleportation via decohered tripartite entangled states}\\[0pt]
\vspace{0.5cm} N. Metwally \\
 Math. Dept., College of Science, University of
Bahrain, Kingdom of Bahrain\\
 Math. Dept., Faculty of science,  University of Aswan, Aswan,
 Egypt.
\end{center}

\begin{abstract}
The entanglement behavior of two classes of multi-qubit system,
GHZ and GHZ like states passing through a generalized amplitude
damping channel is discussed. Despite this channel causes
degradation of the entangled properties and consequently  their
abilities to perform quantum teleportation, one can always improve
the lower values of the entanglement and the fidelity of the
teleportrd state by controlling  on Bell measurements, analyzer
angle and channel's strength. Using GHZ-like state within a
generalized amplitude damping channel is much better than using
the normal GHZ-state, where the decay rate of entanglement and the
fidelity of the teleported states are smaller than those depicted
for GHZ state. 

\end{abstract}

\section{Introduction}
Decoherence represents one of the inevitable phenomena in quantum
information tasks. It is possible to generate maximum entangled
state, but keeping it isolated  from its surrounding is a big
challenge. Therefore, sometimes we are forced to use these
decohered entangled states to achieve some quantum information
tasks. So, investigating the entangled properties of initial
entangled states passing through noise channels are  of a great
importance.

Practically, it is often required to send some parts of the
generated entangled states from the source to remote users
\cite{kimble}. Each transmitted subsystem interacts with its
environment locally and consequently leads to lose some of
entangled properties of the multipartite system \cite{Siomau,Wang,
Paula, Metwally2013}. Consequently, the efficiency of using these
decohered entangled states as quantum channels to perform quantum
teleportation decreases. However, there are some efforts have been
introduced to recover and protect entanglement from degradation
\cite{Mari,Man}. Despite  this degradation, it has been shown that
local environment can enhance the fidelity of quantum
teleportation \cite{Horodecki,Bandy}.

The dynamics of a single and two qubit states in noisy channel has
been studied extensively. The most important channels  are phase
flip, depolarize  and amplitude damping channels
\cite{Nielsen,Nagwa}. Recently, Dontealegre et. al have shown that
the effect of the generalized amplitude damping   channel can be
frozen \cite{Paula,Silva}. Metwally \cite{Metwally2013} has shown
that under the effect of the generalized amplitude channel, the
entanglement of different classes of two qubit systems is stable
and fixed  for larger interval of the channel strength.

It is well known that, quantum teleportation is one of the most
important applications of entanglement. Since the first quantum
teleportation protocol proposed by Bennett et. al \cite{Bennt}, to
teleport a single qubit using an entangled qubit system has been
introduced, there are several versions have been suggested (see
for example \cite{Pati1,Pati2,Gordon}). The possibility of
teleporting an unknown qubit  using tripartite GHZ state is
discussed by Karlsson and Bourennane \cite{Karlsson}. Gorbachev
and Turbilko have introduced  a teleportation protocol to teleport
a two-qubit state by using GHZ state \cite{Gorb}. Another  class
of tripartite entangled state  called W-state has been used to
teleportate an unknown state probabilistically \cite{Shi}. In
2009, Yang et. al have introduced a quantum teleportation scheme
to teleport an unknown  single qubit by using a different class of
GHZ called GHZ-like state \cite{Yang}.

In this work, we investigate the  effect of the generalized
amplitude damping channel on  two classes of tripartite entangled
states: GHZ and GHZ- like states. In this study, we try to answer
the following questions: (i) Is the effect of the generalized
amplitude damping channel on a tripartite states  similar to that
predicted for two qubit systems as shown in \cite{Paula,
Metwally2013}?. In  other words, can one freeze the effect of the
generalized amplitude damping channel when tripartite  state
passes through it. (ii) Can the fidelity of the teleported state
by using these decohered tripartite states can be improved due to
the local interaction as predicted for two qubit systems
\cite{Horodecki,Bandy}?. (iii) Which state is more robust against
this  type of noisy channel; GHZ or GHZ-like state.?

The paper is organized as follows. In Sec. 2, a description of the
suggested model is introduced. The behavior of entanglement is
discussed in Sec. 3. The possibility of using the decohered
entangled states as quantum channels to perform quantum
teleportation is studied in Sec. 4. Finally,  conclusions are
drawn in Sec. 5.

\section{The system}
 It is assumed that a source  generates  states of three qubits  in the GHZ or GHZ-like
 forms as:

\begin{eqnarray}\label{iniS}
 \ket{\psi^{(ini)}}&=& \left\{ \begin{array}{ll}
\ket{\psi_{g}}=\frac{1}{\sqrt{2}}(\alpha\ket{000}+\beta\ket{111}),&  \\
\ket{\psi_{g\ell}}=\frac{1}{2}(c_1\ket{001}+c_2\ket{1010}+c_3\ket{100}+c_4\ket{111}),&  \\
\end{array} \right\}
\end{eqnarray}
where $\alpha^2+\beta^2=1$ and $c_1^2+c_2^2+c_3^2+c_4^2=4$. From
these states we get the maximum entangled of the GHZ state
($\ket{\psi_g}$) and the GHZ-like state ($\ket{\psi_{gl}})$ by
setting $\alpha=\beta=1$ and $c_1=c_2=c_3=c_4=1$ respectively.
 During the transition from the source to the three users
 Alice, Bob and Charlie, the qubits are forced to pass through a
 generalized amplitude damping channel, which is defined by the
 following Kraus operators
 \cite{Nagwa}
 \begin{eqnarray}
 \mathcal{E}_0&=&\frac{\sqrt{p}}{2}\Bigl\{(1+\sqrt{1-\gamma})+(1-\sqrt{1-\gamma})\sigma_z^{(i)}\Bigr\},
 ~\quad\quad
 \mathcal{E}_1=\sqrt{p}\Bigl\{\sigma_x^{(i)}+i\sigma_y^{(i)}\Bigr\},
 \nonumber\\
 \mathcal{E}_2&=&\frac{\sqrt{1-p}}{2}\Bigl\{(1+\sqrt{1-\gamma})-(1-\sqrt{1-\gamma})\sigma_z^{(i)}\Bigr\},
 \quad
 \mathcal{E}_3=\sqrt{1-p}\sqrt{\gamma}\Bigl\{\sigma_x^{(i)}-\sigma_y^{(i)}\bigr\}
 \end{eqnarray}
where $\sigma_j^{(i)}, j=x,y,z$ and $i=1,2,3$ are the Pauli
operators for the three qubits, respectively, $p$ and $\gamma$ are
the strength and damping parameters of the channel. If we assume
that all the three qubits are  passing through this noise channel,
then  the final state  can be written as:
\begin{equation}
\rho^{(f)}_{k}=\sum_{n=0}^{n=3}\mathcal{E}_n\rho_{k}^{(ini)}\mathcal{E}_n^{\dagger},
\end{equation}
where $\rho_k^{(ini)}=\ket{\psi_k^{(ini)}}\bra{\psi^{(ini)}_k}$ is
the initial state of the travelling  state through the generalized
amplitude damping channel, $k=g$ or $g\ell$ and $\rho_k^{(f)}$ is
the final state.

 In this subsection, we find
 the final state of travelling qubits in the noise
channel  (2) analytically. If we assume that, the  system is
initially prepared in the GHZ state as defined in Eq.(1), then by
using Eq.(2), the final state (3) can be written explicitly as by
\begin{equation}
\rho_g^{(f)}=\mathcal{A}_1\ket{000}\bra{000}+\mathcal{A}_2\ket{000}\bra{111}+\mathcal{A}_3\ket{111}\bra{000}
+\mathcal{A}_4\ket{111}\bra{111},
\end{equation}
where,
\begin{eqnarray}
\mathcal{A}_1&=&\frac{\alpha^2}{2}\Bigl(p^2+(1-p)^3(1-\gamma)^3\Bigr)
,\quad
\mathcal{A}_2=\frac{1}{2}(1-\gamma)^{3/2}\Bigl(p^2\alpha\beta^*+\alpha^*\beta(1-p)^3\Bigr),
\nonumber\\
\mathcal{A}_3&=&\frac{1}{2}(1-\gamma)^{3/2}\Bigl(p^2\alpha^*\beta+\alpha\beta^*(1-p)^3\Bigr),\quad
\mathcal{A}_4=\frac{\beta^2}{2}\Bigl(p^2(1-\gamma)^3+(1-p)^3\Bigr),
\end{eqnarray}
Similarly, if the travelling state is initially prepared in the
GHZ-like state $\rho_{g\ell}$ and passes through the generalized
amplitude damping  channel (2), then the final state
$\rho^{(f)}_{g\ell}$ is given by
\begin{eqnarray}
\rho_{g\ell}^{(f)}&=&\ket{001}\Bigl\{\mathcal{B}_1\bra{001}+\mathcal{B}_2\bra{010}+\mathcal{B}_3\bra{100}+
\mathcal{B}_4\bra{111}\Bigr\}
\nonumber\\
&&+\ket{010}\Bigl\{\mathcal{B}_5\bra{001}+\mathcal{B}_6\bra{010}+\mathcal{B}_7\bra{100}+
\mathcal{B}_8\bra{111}\Bigr\}
\nonumber\\
&&+\ket{100}\Bigl\{\mathcal{B}_9\bra{001}+\mathcal{B}_{10}\bra{010}+\mathcal{B}_{11}\bra{100}+
\mathcal{B}_{12}\bra{111}\Bigr\}
\nonumber\\
&&+\ket{111}\Bigl\{\mathcal{B}_{13}\bra{001}+\mathcal{B}_{14}\bra{010}+\mathcal{B}_{15}\bra{100}+
\mathcal{B}_{16}\bra{111}\Bigr\},
\end{eqnarray}
where,
\begin{eqnarray}
\mathcal{B}_1&=&|c_3|^2\kappa_1,\quad
\mathcal{B}_2=c_3c_1^*\kappa_1, \quad
\mathcal{B}_3=c_3c_2^*\kappa_1,\quad
\mathcal{B}_4=c_3c_4^*\kappa_2+\frac{p^3}{4}c_4c_3^*,
 \nonumber\\
\mathcal{B}_5&=&c_1c_3^*\kappa_1, \quad
\mathcal{B}_6=|c_1|^2\kappa_1, \quad
\mathcal{B}_7=c_1c_2^*\kappa_1,\quad
\mathcal{B}_8=c_1c_4^*\kappa_2,
\nonumber\\
\mathcal{B}_9&=&c_2c_3^*\kappa_1, \quad
\mathcal{B}_{10}=c_2c_1^*\kappa_1, \quad
\mathcal{B}_{11}=|c_2|^2\kappa_1,\quad
\mathcal{B}_{12}=c_2c_4^*\kappa_2,
 \nonumber\\
\mathcal{B}_{13}&=&c_4c_1^*\kappa_2, \quad
\mathcal{B}_{14}=c_4c_1^*\kappa_2, \quad
\mathcal{B}_{15}=c_4c_2^*\kappa_2,\quad
\mathcal{B}_{12}=|c_4|^2\kappa_3,
\end{eqnarray}
with,
\begin{eqnarray}
\kappa_1&=&\frac{1-\gamma}{4}(p\sqrt{p}+(1-p)^{3/2}),
\nonumber\\
\kappa_2&=&\frac{1-\gamma}{4}(p\sqrt{p}(1-\gamma)+(1-p)^{3/2}),
\nonumber\\
\kappa_3&=&\frac{1-\gamma}{4}(p\sqrt{p}(1-\gamma)^3+(1-p)^{3/2}).
\end{eqnarray}
Since  the final states of GHZ and GHZ-like states have been
obtained, we can quantify the survival amount of entanglement.
Also, the possibility of using these decohered states as quantum
channel to perform quantum teleportation will be discussed in Sec.
(4).

\section{Entanglement}
In this section, we quantify the  survival amount of entanglement
which is contained in the travelling state through the noisy
channel. In this context, we use the tripartite negativity as a
measure of entanglement. This measure states that, if $\rho_{abc}$
represents a tripartite state, then the negativity is defined as,
\begin{equation}
\mathcal{N}(\rho_{abc})=(\mathcal{N}_{a-bc}\mathcal{N}_{b-ac}\mathcal{N}_{c-ab})^{\frac{1}{3}},
\end{equation}
where
$\mathcal{N}_{i-jk}=-2\sum_{\ell}{\lambda_{\ell}(\rho_{ijk}^{T_i})}$,
$\lambda_{\ell}$ are the negative eigenvalues of the partial
transpose of the state $\rho_{ijk}$ with respect to the qubit
$"i"$ \cite{Carlos}.
\begin{figure}[t!]
  \begin{center}
    \includegraphics[width=20pc,height=15pc]{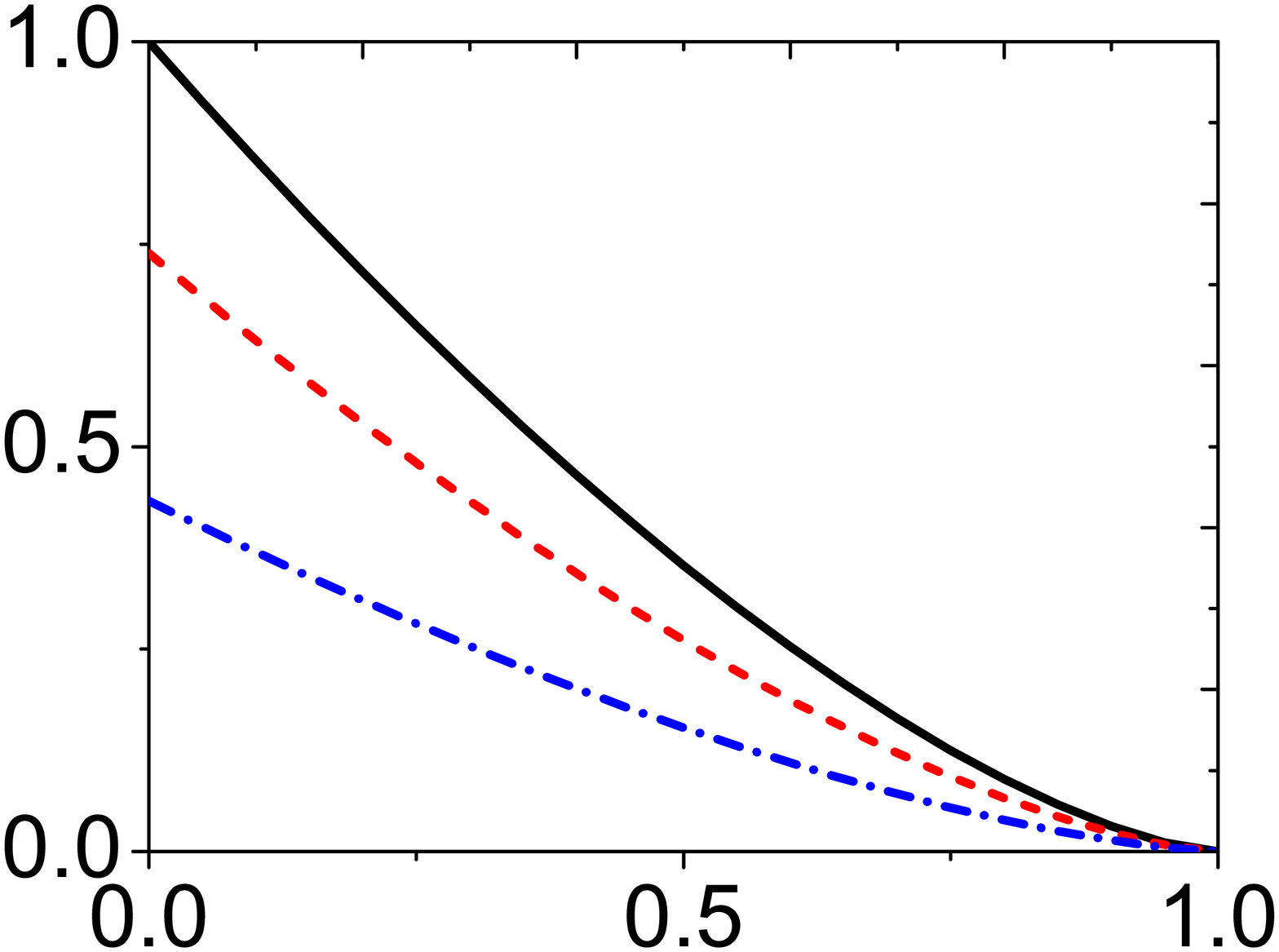}
   \put(-260,90){$\mathcal{N}(\rho_{g}^{(f)})$}
   \put(-125,-1){\Large$\gamma$}
   \put(-60,145){ $(a)$}
  \includegraphics[width=20pc,height=15pc]{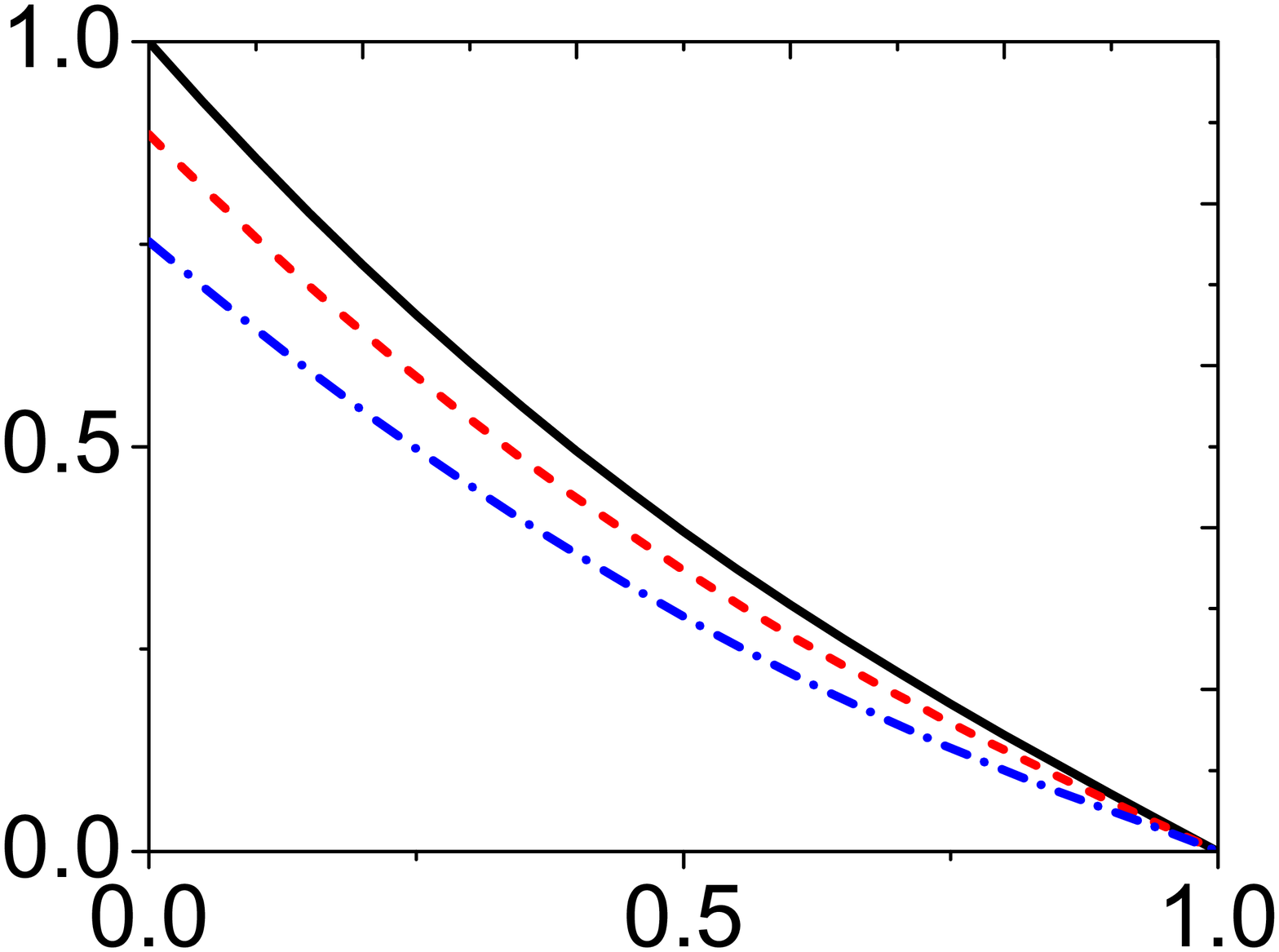}
      \put(-258,90){$\mathcal{N}(\rho_{g\ell}^{(f)})$}
   \put(-125,-1){\Large$\gamma$}
   \put(-60,145){ $(b)$}\\
   \includegraphics[width=20pc,height=15pc]{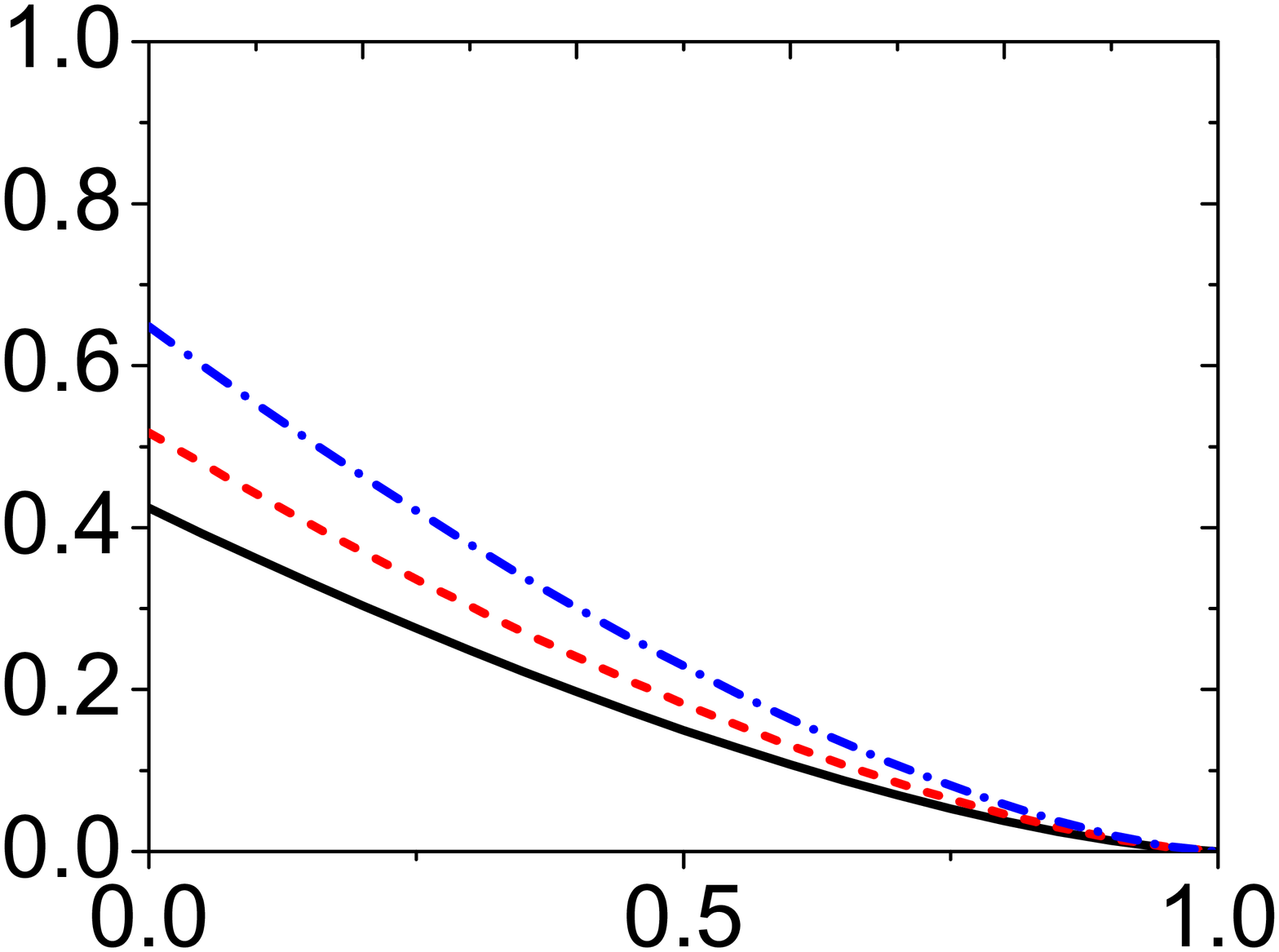}
   \put(-260,90){$\mathcal{N}(\rho_{g}^{(f)})$}
   \put(-125,-1){\Large$\gamma$}
   \put(-60,145){ $(c)$}
  \includegraphics[width=20pc,height=15pc]{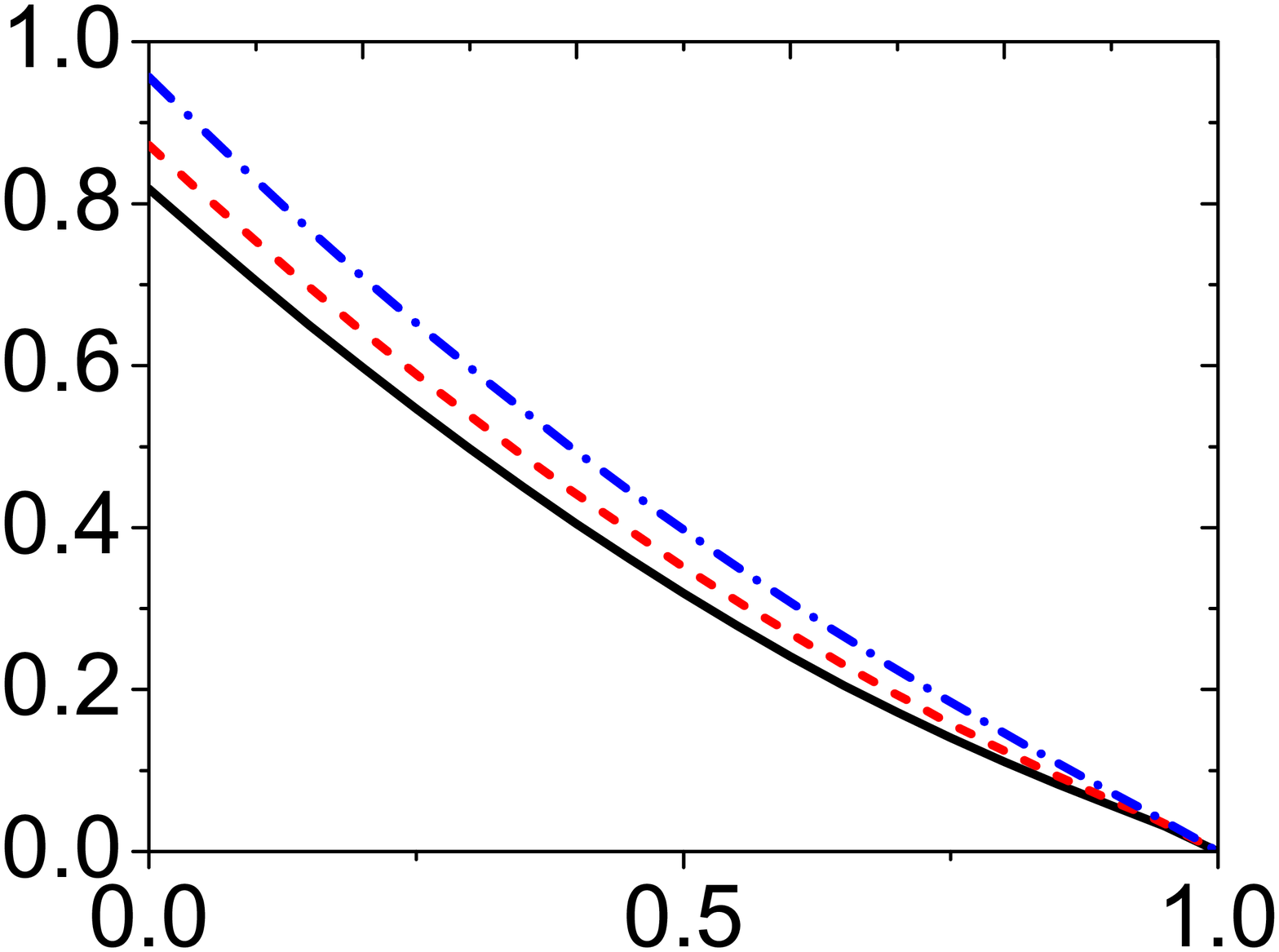}
      \put(-255,90){$\mathcal{N}(\rho_{g\ell}^{(f)})$}
 \put(-125,-1){\Large$\gamma$}
 \put(-60,145){ $(d)$}
     \caption{The entanglement $\mathcal{N}_{g}(\rho^{(f)}) (Figs.(a,c))$
      and  $\mathcal{N}_{g\ell}(\rho^{(f)}) (Figs.(b,d))$
     for system initially prepared in GHZ and  GHZ-like
     state, respectively. For figures (a,b), $p=0, 0.1$ and
    $0.3$  for the solid, dash, and dash-dot curves, respectively,
   while for  figures (c,d), $p=0.6, 0.7$ and $0.8 $  for the solid, dash, and dash-dot curves, respectively.}
       \end{center}
\end{figure}

In Fig.(1),  the survival amount of entanglement between the three
qubits who initially share a GHZ  or  GHZ-like state is displayed.
In Fig.(1a), the entanglement  behavior of a system  initially
prepared in GHZ state with $\alpha=\beta=1$ passing through a
generalized amplitude damping channel, is shown.  The general
behavior shows that, $\mathcal{N}_g(\rho^{(f)})$ decays as
$\gamma$ increases to vanish completely at $\gamma=1$. On the
other hand, as  the strength of the channel $p$ increases, the
upper bounds of
 entanglement decrease.
In Fig.(1b), it is  assumed that GHZ-like state is initially
prepared with $c_i=1, i=1,2,3,4$. It is clear that, for
$p=\gamma=0$, the entanglement is maximum i.e.,
$\mathcal{N}_{g\ell}(\rho^{(f)})=1$. However, as $\gamma$
increases further, the entanglement decays gradually to vanish
completely at $\gamma=1$. For small values of the channel
strength, the entanglement decays gradually, while the rate of
decay increases as one increases the channel strength
$p\in[0,0.5]$.

The behavior of entanglement for larger values of the channel's
strength, $p$ is displayed in Figs.(1c) for GHZ  and (1d) for
GHZ-lLike state. The behavior of Entanglement is similar to that
depicted in Figs.(1a) and (1b), namely the entanglement decays as
$\gamma$ increases. However as one increases the channel strength
$p$, the entanglement increases and the decay rate decreases.
Comparing these two figures,  one can see that the decay rate for
the GHZ state is larger than that displayed for GHZ-like state.
\begin{figure}
  \begin{center}
\includegraphics[width=21pc,height=15pc]{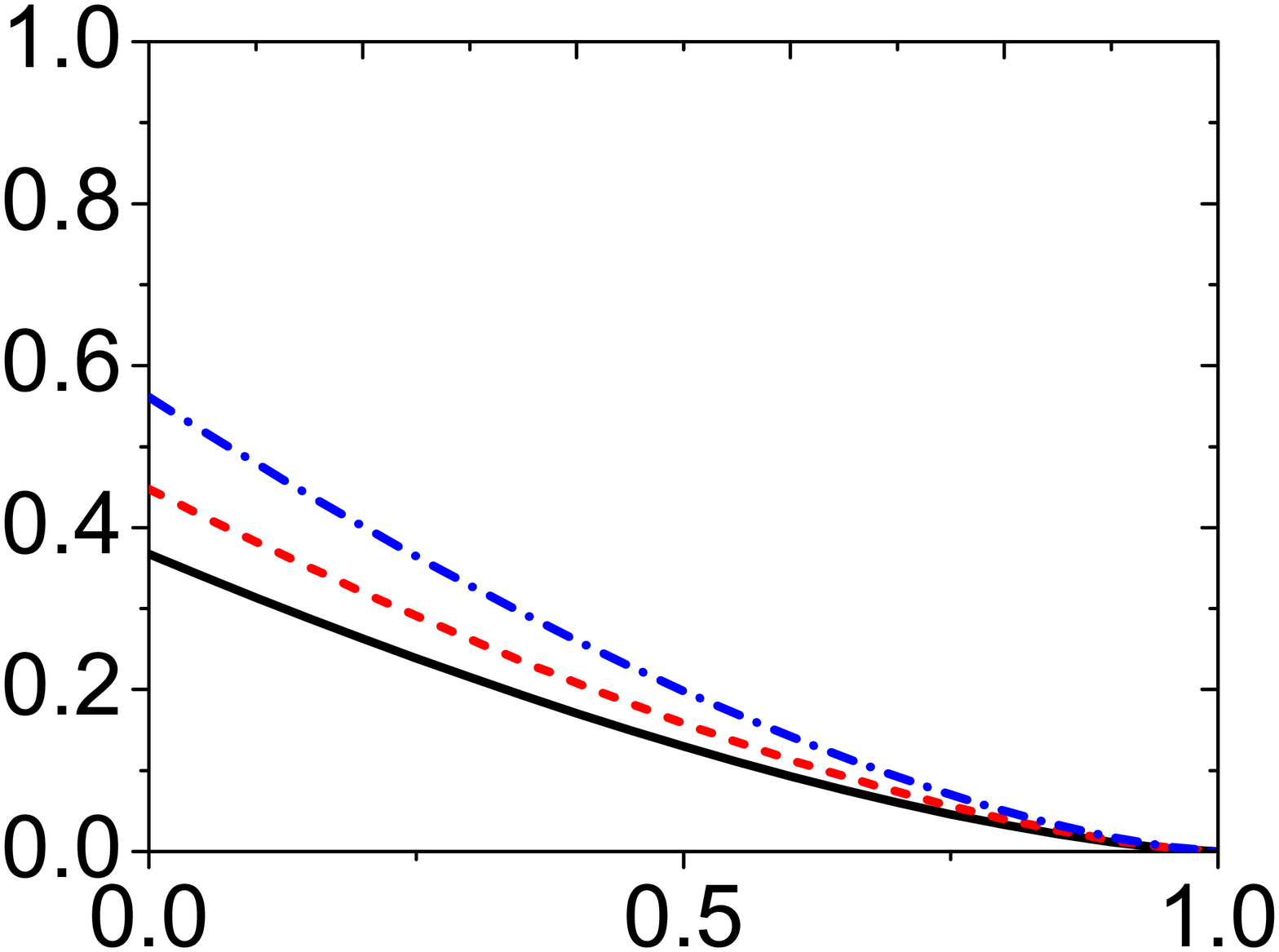}\quad
 \put(-260,90){$\mathcal{N}(\rho_{g}^{(f)})$}
  \put(-145,-1){\Large$\gamma$}
  \put(-60,145){ $(a)$}
  \includegraphics[width=21pc,height=15pc]{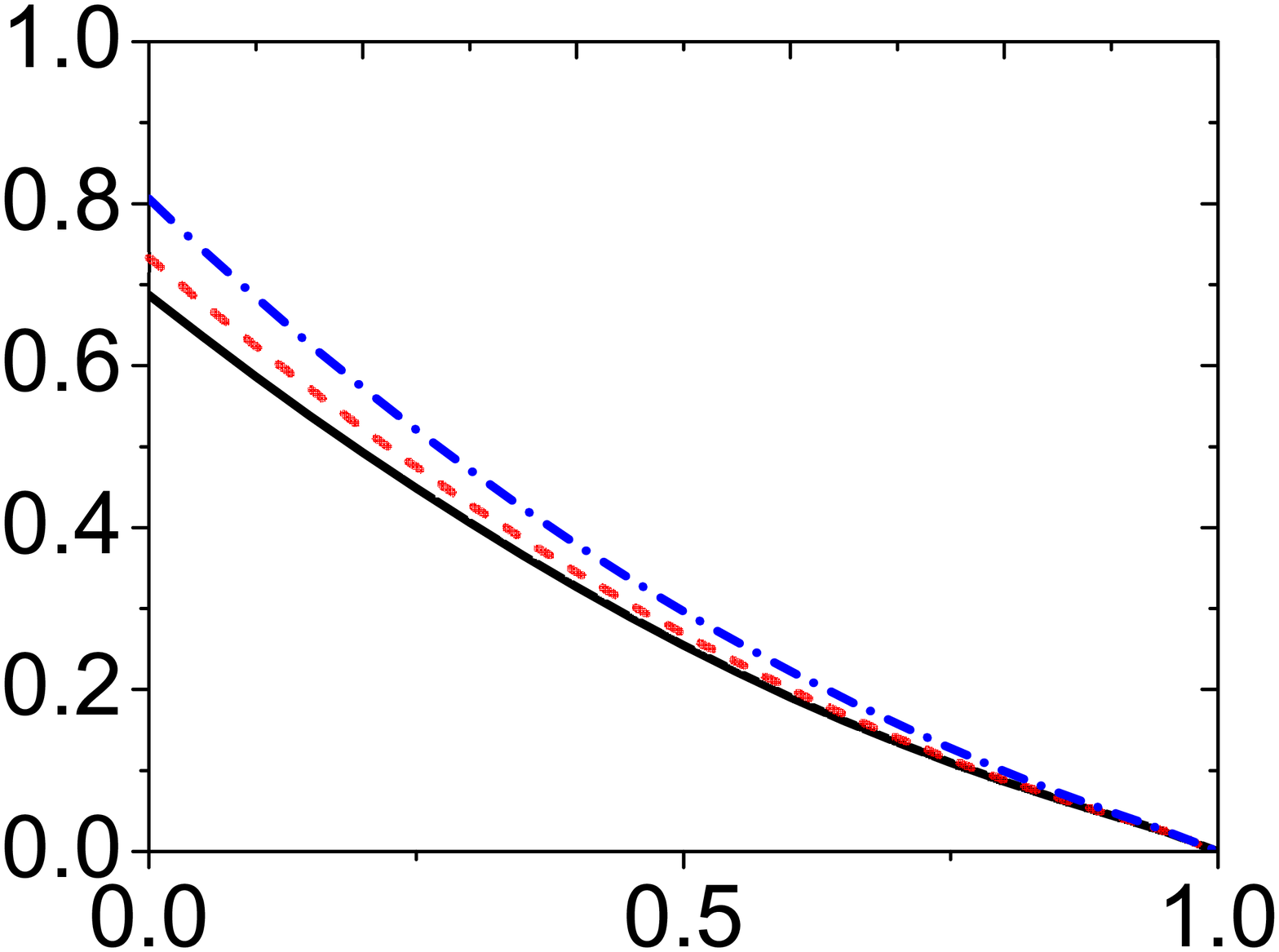}
  \put(-260,90){$\mathcal{N}(\rho_{g\ell}^{(f)})$}
 \put(-145,-1){\Large$\gamma$}
 \put(-60,145){ $(b)$}
     \caption{The solid, dash and dash-dot curves,represent the entanglement for  $p=0.6,0.7$ and $0.8 $
   respectively,where the systems are prepared initially in
  (a) GHZ state  with  $\alpha=0.5$ and $\beta=\sqrt{3}/2$, and
    (b)GHZ-like state with  $c_1=0.8,c_2=0.7, c_3=0.6$ }
       \end{center}
\end{figure}

In reality, it is difficult to keep the generated  maximum
entangled state isolated from  its surroundings and consequently
it turns into partial entangled states. Therefore, it is important
to investigate the behavior of these  partial entangled states in
the presence of this noise. For this aim, we consider  Fig.(2),
where it is assumed that, the users share non-maximally entangled
states.  Fig.(2a), displays the decohered GHZ state which is
initially prepared with $\alpha=0.5$ and $\beta=\sqrt{3}/2$. The
general behavior is similar to that depicted in Fig.(1) i.e., the
entanglement decays as $\gamma$ increases. However, the upper
bounds  of entanglements are always smaller than those  shown for
maximum entangled state (see Fig.(1c)). In Fig.(2a), we consider
that the users share a non-maximum entangled GHZ-like state
defined by $c_1=0.8, c_2=0.7$ and $c_3=0.6$, where we consider a
larger values of the channel strength. The general behavior is
similar to that predicated in Fig.(1), but as $p$ increases the
decay rate decreases and consequently the upper bounds of
entanglement are  larger.

\section{Teleportation}
In this section, we investigate the effect of the generalized
amplitude damping channel on the fidelity of the teleported state.
We consider the decohered GHZ  and GHZ-like states as quantum
channel between the three users to perform the teleportation
protocol. In this current investigation, we assume that the three
users co-operate to achieve this protocol. Assume that Alice is
given unknown information coded in the single qubit
$\rho_u=\ket{\psi_ui}\bra{\psi_u}$, where
$\ket{\psi_u}=\mu\ket{0}+\nu\ket{1}, |\mu|^2+|\nu|^2=1$. The
initial system between the three users is given by
$\rho_s=\rho_u\otimes\rho_{g}^{(f)}$ or
$\rho_s=\rho_u\otimes\rho_{g\ell}^{(f)}$. Now, Alice has two
qubits: her own qubit and the unknown qubit, while Bob and Charlie
have the second and the third qubits respectively. To perform the
teleportation protocol, the users follow the following steps:

\begin{enumerate}
\item  Alice performs Bell measurements (BM), i.e.
$\rho_{\phi^\pm}=\ket{\phi^\pm}\bra{\phi^{\pm}}$,
$\rho_{\psi^{\pm}}=\ket{\psi_{\pm}}\bra{\psi_{\pm}}$,
$\ket{\phi^{\pm}}=\frac{1}{\sqrt{2}}(\ket{00}\pm\ket{11}$,
$\ket{\psi^{\pm}}=\frac{1}{\sqrt{2}}(\ket{00}\pm\ket{11}$  on the
first two qubits; (here qubit and the unknown qubit, and Charlie
makes her  measurements on the basis either $"0"$ or $"1"$.

\item Alice and Charlie send their measures to Bob, who will do
the appreciated operations  to get the decohered teleported
information.
\end{enumerate}

\subsection{ Decohered GHZ state as quantum channel}
 Let us first, assume that the initial state of the system is
$\rho_s=\rho_u\otimes\rho_{g}^{(f)}$, i.e., the users will use the
decohered GHZ state as quantum channel. Assume that, Alice
performs here measurements  by using Bell state analyzers on
qubits $"u"$ and $"1"$ \cite{Karlsson}. If the Bell state analyzer
gives $\ket{\phi^+}_{u1}$, or $\ket{\phi^-}_{u1}$ then, the state
of the Charlie and Bob will be  projected into
\begin{equation}\label{CBPP}
\rho^{\phi^{\pm}}_{CB}=\kappa_{00}\ket{00}\bra{00}\pm\kappa_{01}\ket{00}\bra{11}\pm\kappa_{10}\ket{11}\bra{00}+\kappa_{11}\ket{11}\bra{11},
\end{equation}
where,
\begin{eqnarray}
\kappa_{00}&=&A_1(\mu^2+\mu^*\nu+\mu^*\nu^*+\nu^2), \quad
\kappa_{01}=A_2(\mu^2-\mu^*\nu+\mu^*\nu^*-\nu^2),
 \nonumber\\
\kappa_{10}&=&A_3(\mu^2+\mu^*\nu-\mu^*\nu^*-\nu^2), \quad
\kappa_{11}=A_4(\mu^2-\mu^*\nu-\mu^*\nu^*+\nu^2).
\end{eqnarray}
 To complete the protocol, Charlie,  uses
a spin-state analyzer with two outcomes
$\ket{x_1}=\frac{1}{2}(\sin\theta\ket{1}+\cos\theta\ket{0})$ and
$\ket{x_2}=\frac{1}{2}(\cos\theta\ket{1}-\sin\theta\ket{0})$,
where $\theta$ is the analyzer angle, to measure her qubit
\cite{Karlsson}. However, if Charlie measures $\ket{x_1}$, or
$\ket{x_2}$ then  Bob  will get the final state with a fidelity
given by,
\begin{eqnarray}
{\mathcal{F}_{g}}^{(x_1)}_{\phi^+}&=&
  \mu^2\kappa_{00}\cos^2\theta+\frac{1}{2}\sin2\theta(\mu\nu^*\kappa_{01}+
  \mu^*\nu\kappa_{10})+\nu^2\kappa_{11}\sin^2\theta,
  \nonumber\\
  {\mathcal{F}_{g}}^{(x_2)}_{\phi^+}&=&
  \mu^2\kappa_{00}\sin^2\theta+\frac{1}{2}\sin2\theta(\mu\nu^*\kappa_{01}+
  \mu^*\nu\kappa_{10})+\nu^2\kappa_{11}\cos^2\theta.
  \end{eqnarray}
 On the other hand, if the Bell analyzer  read out is
$\ket{\psi}^{(\pm)}_{u1}$, then the state  between Charlie and Bob
is projected into,
\begin{equation}
\rho^{\psi^\pm}_{CB}=\kappa_{11}\ket{00}\bra{00}\pm\kappa_{10}\ket{00}\bra{11}\pm\kappa_{01}\ket{11}\bra{00}
+\kappa_{00}\ket{11}\bra{11},
\end{equation}
where $\kappa_{ij}, ij=00,01,10$ and $11$ are given by Eq.(11).
The details of measurements and operations which can be done by
the users are shown in Table (1).

\begin{table}
  \centering
  \begin{tabular}{|c|c|c|c|c|}
\hline
   Alice& Chirile & Bob &Fidelity\\
  \hline\hline
  $\rho_{\phi^+}$ & $\ket{x_1}$ & $I$&${\mathcal{F}_{g}}^{(x_1)}_{\phi^+}$\\
 &$\ket{x_2}$&$S_z$&${\mathcal{F}_{g}}^{(x_2)}_{\phi^+}$\\
  \hline
 $\rho_{\phi^-}$ & $\ket{x_1}$ & $S_z$ & ${\mathcal{F}_{g}}^{(x_1)}_{\phi^-}={\mathcal{F}_{g}}^{(x_1)}_{\phi^+}$\\
 &$\ket{x_2}$&$I$&${\mathcal{F}_{g}}^{(x_2)}_{\phi^-}={\mathcal{F}_{g}}^{(x_2)}_{\phi^+}$\\
 \hline
  $\rho_{\psi^+}$ & $\ket{x_1}$ & $I$& ${\mathcal{F}_{g}}^{(x_1)}_{\psi^+}=
  {\mathcal{F}_{g}}^{(x_1)}_{\phi^+}$\\
  &$\ket{x_2}$&$S_z$&${\mathcal{F}_{g}}^{(x_2)}_{\psi^+}={\mathcal{F}_{g}}^{(x_2)}_{\phi^+}$\\
  \hline
    $\rho_{\psi^-}$&$\ket{x_1}$ & $S_z$&${\mathcal{F}_{g}}^{(x_1)}_{\psi^-}={\mathcal{F}_{g}}^{(x_1)}_{\phi^+}$\\
  &$\ket{x_2}$&$I$&${\mathcal{F}_{g}}^{(x_2)}_{\psi^-}={\mathcal{F}_{g}}^{(x_2)}_{\phi^+}$\\
  \hline
\end{tabular}
 \caption{ Teleportation protocol via decohered  GHZ state as quantum channel}
\end{table}

\begin{figure}
  \begin{center}
\includegraphics[width=21pc,height=15pc]{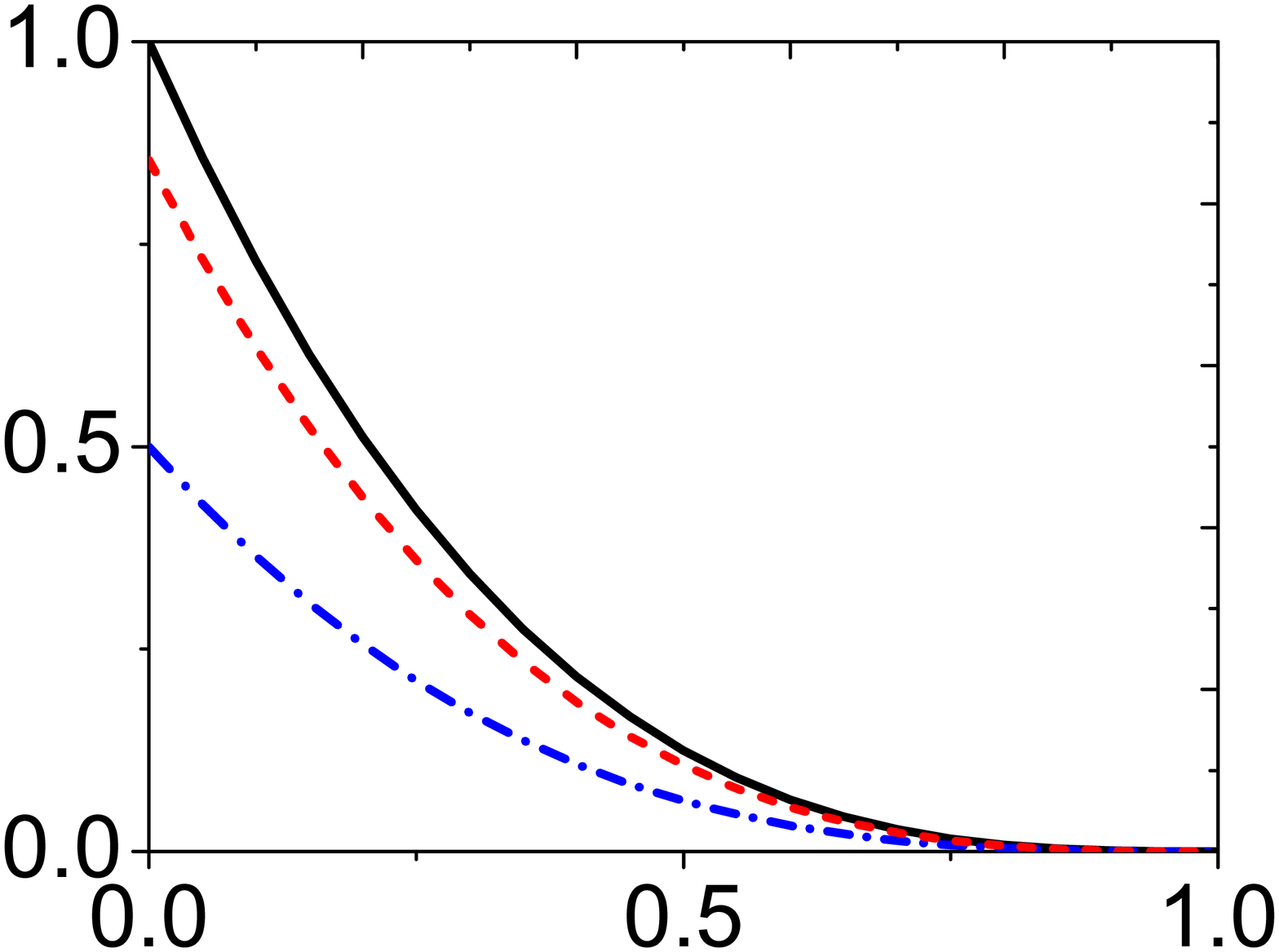}\quad
 \put(-260,90){$\mathcal{F}^{(x_1)}_{g\phi^+}$}
  \put(-125,-1){\Large$\gamma$}
   \put(-60,145){ $(a)$}
  \includegraphics[width=21pc,height=15pc]{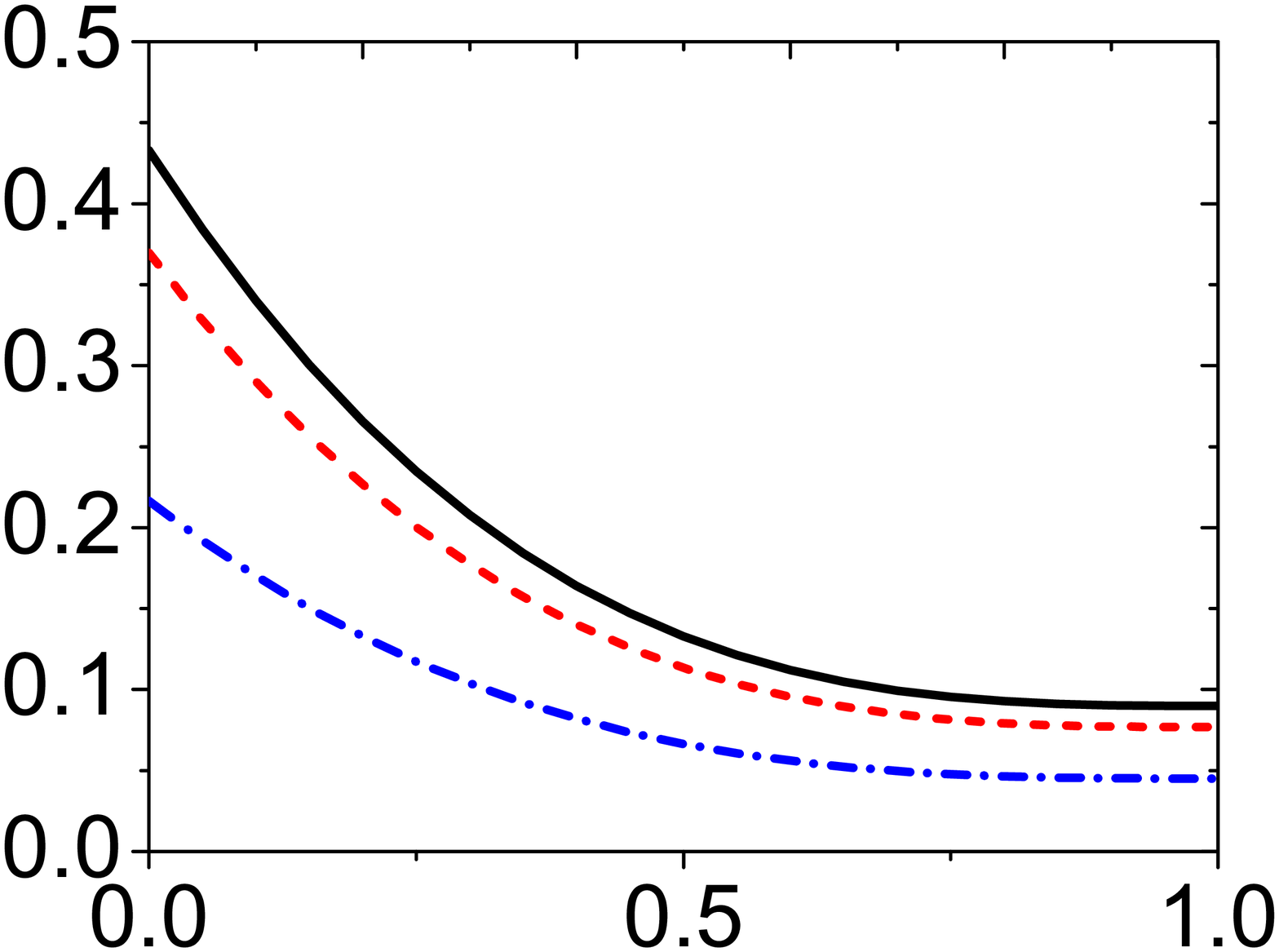}
  \put(-260,90){$\mathcal{F}^{(x_1)}_{g\phi^+}$}
 \put(-125,-1){\Large$\gamma$}
  \put(-60,145){ $(b)$}
     \caption{ The fidelity of the teleported state, with the
     channel strength,
   showing  solid, dash and dash-dot curves
      for $\theta=0,\pi/8$ and $\pi/4$, respectively for  (a) $p=0$ and
     (b) $p=0.3$.}
       \end{center}
\end{figure}

The behavior of the fidelity $\mathcal{F}^{(x_1)}_{g\phi^+}$ of
the teleported state by using the decohered GHZ-state (4) as
quantum channel is shown in Fig.(3), where different values of the
analyzer angle are considered within and without the channel
strength, $p$. As it is  described in Fig.(3a), the fidelity
decreases as the channel damping parameter $\gamma$ increases. For
small values of analyzer angle ($\theta=0$) and $\gamma=0$, the
initial fidelity is maximum ($\mathcal{F}^{(x_1)}_{g\phi^+}=1)$.
However, for larger values of the channel damping parameter, the
fidelity decays smoothly to vanish completely at $\gamma=1$. The
initial fidelity decreases as one increases the analyzer angle as
 depicted for $\theta=\pi/8$ and $\pi/4$, respectively. Fig.(3b)
displays the effect of the channel strength $p$ on the fidelity of
the teleported state for different values of analyzer angle, where
we set $p=0.3$. It is clear that, the initial fidelity i.e., at
$\gamma=0$, is  very small comparing to that displayed in
Fig.(3a). On the other hand, the decay rate of the fidelity is
smaller than that depicted  for zero value of the channel's
strength (see Fig.(3a)). Moreover, the long-lived fidelity is
depicted for larger values of the channel damping parameter
$(\gamma>0.5)$, where the fidelity is almost constant.

\begin{figure}[t!]
  \begin{center}
    \includegraphics[width=19pc,height=15pc]{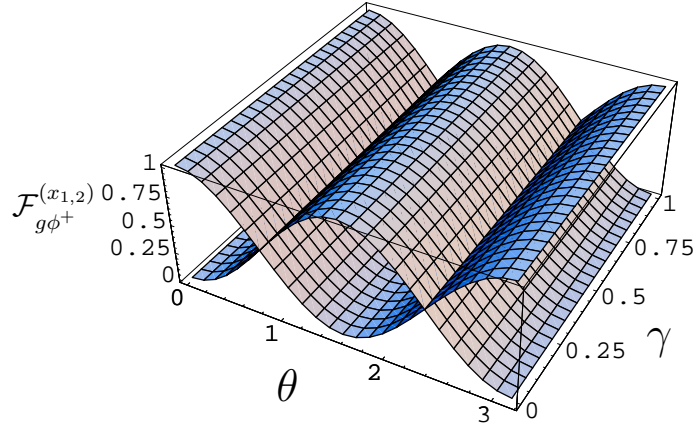}
  \put(-260,90){$\mathcal{F}^{(x_{1,2})}_{g\phi^+}$}
  \put(-160,20){\Large$\theta$}
  \put(-20,40){\Large$\gamma$}
       \caption{ The fidelities  $\mathcal{F}^{(x_{1,2})}_{g\phi^+}$ of the teleported state
      with $\mu=\nu=1/\sqrt{2}$, where the channel's strength $p=0$.}
       \end{center}
\end{figure}

\begin{figure}
  \begin{center}
   \includegraphics[width=21pc,height=15pc]{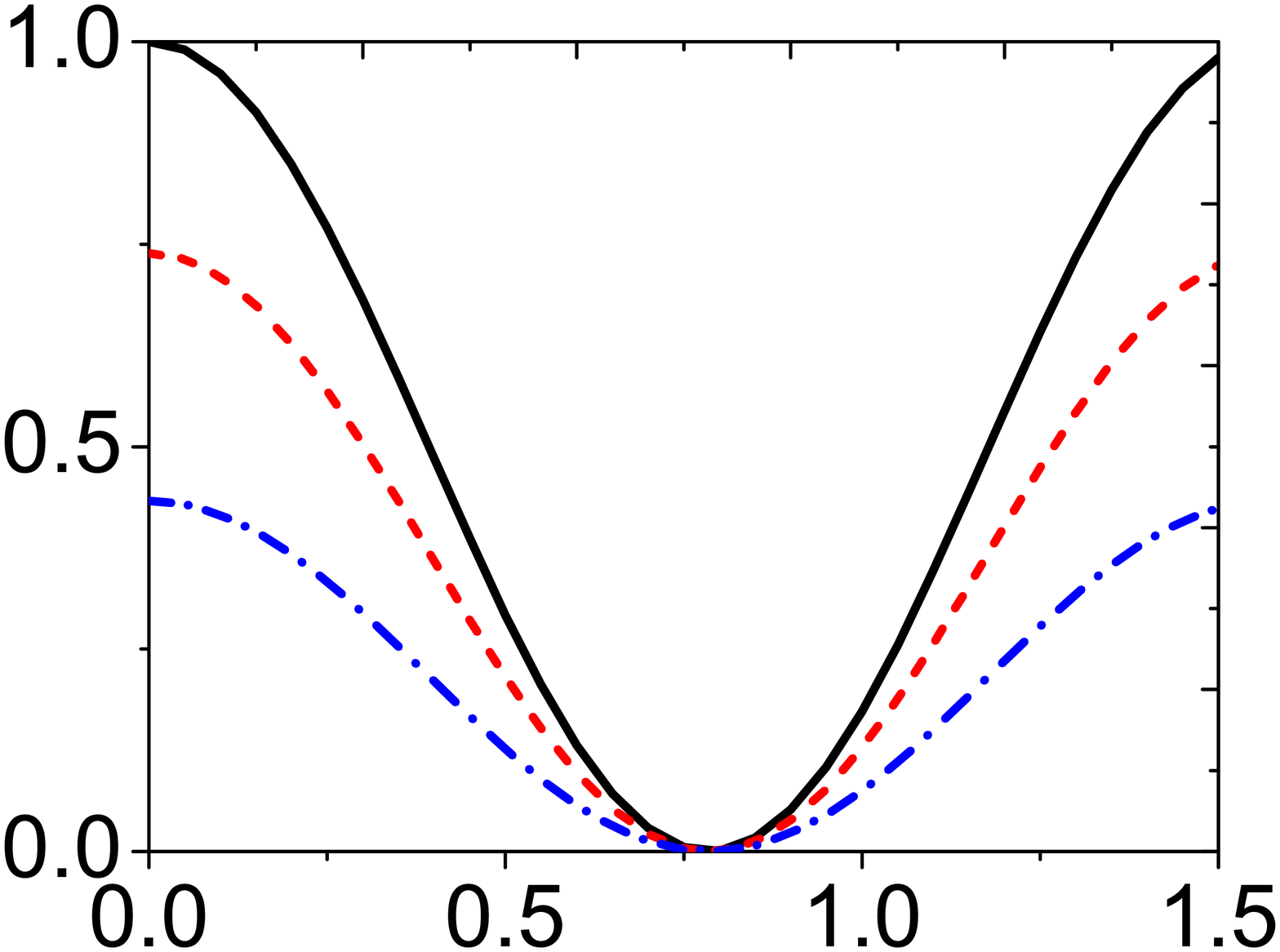}\quad
 \put(-260,90){$\mathcal{F}^{(x_1)}_{g\phi^+}$}
  \put(-125,-1){\Large$\theta$}
   \put(-70,145){ $(a)$}
  \includegraphics[width=21pc,height=15pc]{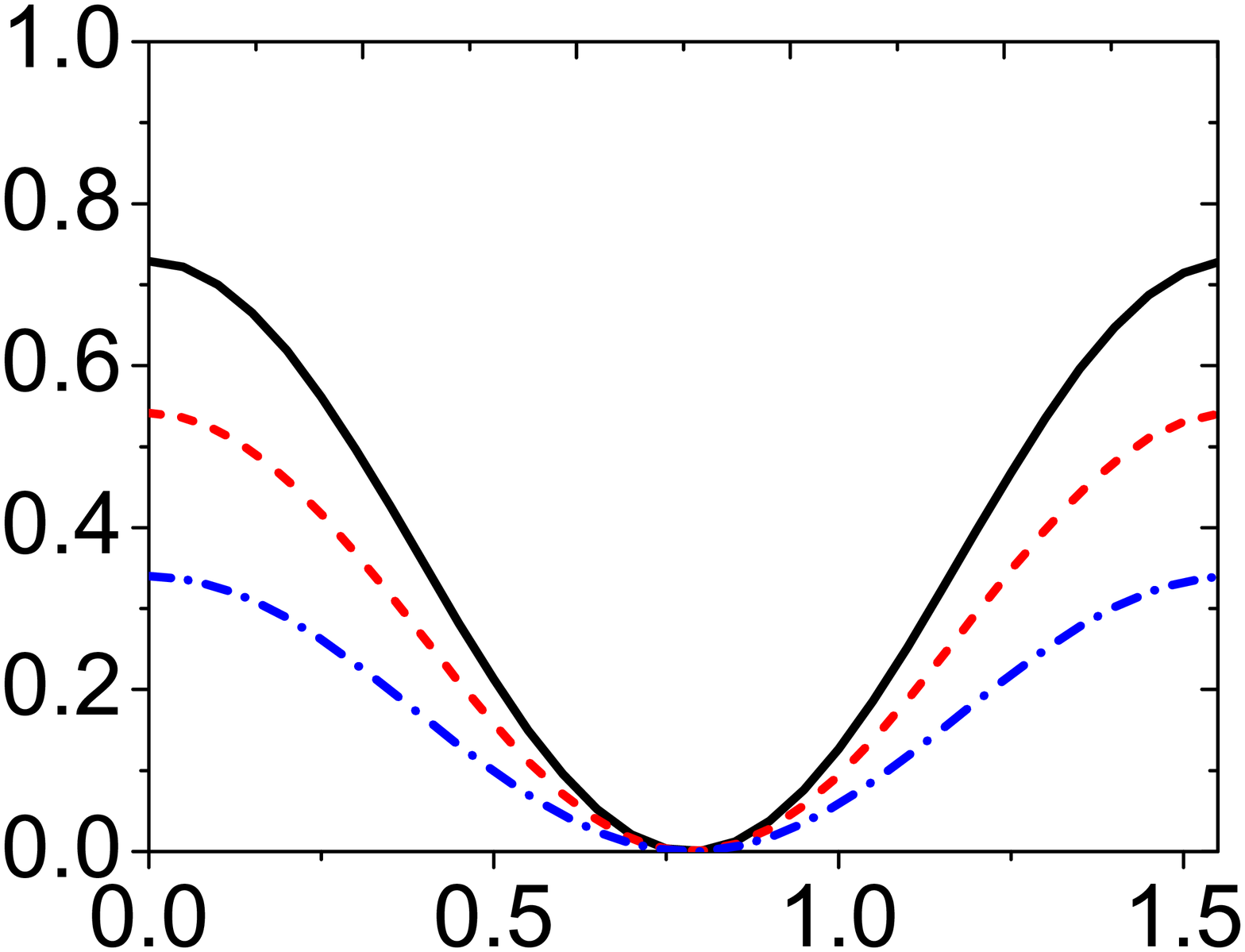}
  \put(-260,90){$\mathcal{F}^{(x_1)}_{g\phi^+}$}
 \put(-125,-1){\Large$\theta$}
  \put(-60,145){ $(b)$}
     \caption{ The fidelity of the teleported state
      as a function of $\theta$. The solid, dash and dash-dot curves
      for$p=0,0.1,0.3$, respectively  for(a) $\gamma=0 $ and
     (b) $\gamma=0.1$.}
       \end{center}
\end{figure}

Fig.(4), shows the behavior of the fidelities
$\mathcal{F}^{(x_1)}_{g\phi^+}$ and
$\mathcal{F}^{(x_2)}_{g\phi^+}$, namely when Charlie measures
$\ket{x_1}$ and $\ket{x_2}$, respectively as functions in the
analyzer angle $\theta$. It is clear that, at $\gamma=0$,
$\mathcal{F}^{(x_1)}_{g\phi^+}$ is maximum, while
$\mathcal{F}^{(x_1)}_{g\phi^+}=0$ i.e., is minimum. As $\theta$
increases further, the fidelity, $\mathcal{F}^{(x_1)}_{g\phi^+}$
decreases while $\mathcal{F}^{(x_2)}_{g\phi^+}$ increases. However
at $\theta=\pi/2$, the situation of the behavior of the two
fidelities is changing. This figure shows that, one can use the
analyzer angle as a control parameter, where if Charlie decides to
measure $\ket{x_1}$ or $\ket{x_2}$, then by adjusting the analyzer
angle, the final state can be obtained with a maximum fidelity.

In Fig.(5), we investigate the behavior of the fidelity as a
function of the analyzer angle, $\theta$ for different values of
the channel strength. Fig.(5a) is devoted to study the behavior of
$\mathcal{F}^{(x_1)}_{g\phi^+}$ with zero damping channel i.e., we
set $\gamma=0$. As shown in Fig.(3a), the initial fidelity
decreases as the analyzer angle $\theta$ increases. It is clear
that, the fidelity decreases to   vanish completely at
$\theta=\pi/2$ and increases again to reach its maximum value at
$\theta=\pi$. The upper bound of the fidelity of the teleported
state decreases as the analyzer angle increases which is in
agreement with the work of Karlsson and Bourennane
\cite{Karlsson}. As one increases the value of the channel damping
parameter $\gamma$, the upper bounds of the fidelity decreases as
shown in Fig.(4b), where we set $\gamma=0.1$.

\subsection{ GHZ-like state as quantum channel}

Now, we assume that the users share a  decohered GHZ-like state
(6), this means that the system is given by
$\rho_s=\rho_u\otimes\rho_{g\ell}^{(f)}$. They use the same
protocol  described above. However depending on the results of
Alice and Charlie, Bob performs the adequate operation to get the
teleported state.  For example, If Alice measures $\rho_{\phi^+}$
and Charlie measures $"1"$, then Bob will do nothing and the
fidelity of the teleported state is given by
${\mathcal{F}_{g\ell}}^{(1)}_{\phi^+}$ as shown in Table (2).

\begin{figure}[t!]
  \begin{center}
\includegraphics[width=21pc,height=15pc]{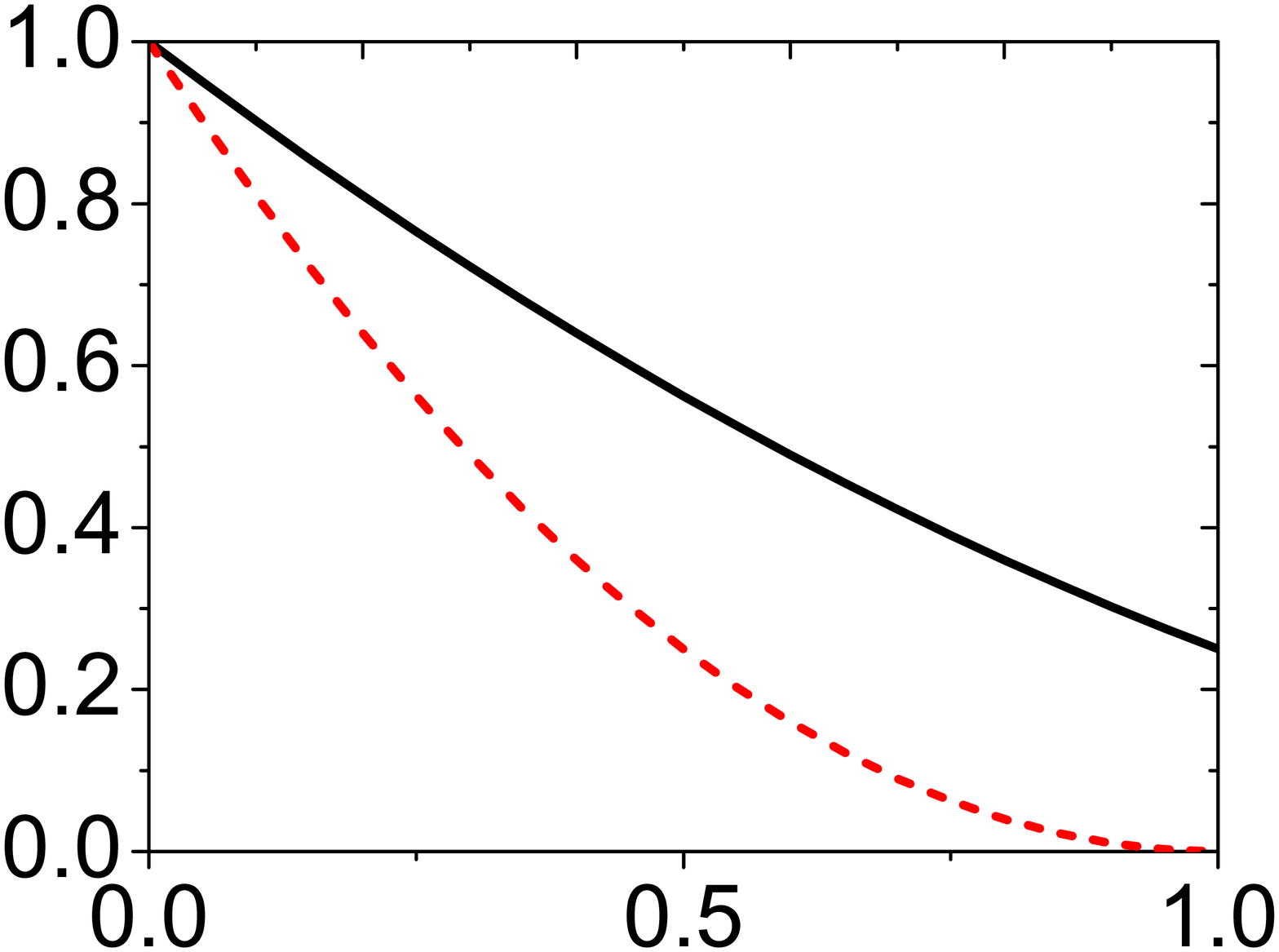}\quad
 \put(-270,90){$\mathcal{F}(\rho_{g\ell}^{(f)})$}
  \put(-125,-1){\Large$\gamma$}
   \put(-60,145){ $(a)$}
  \includegraphics[width=21pc,height=15pc]{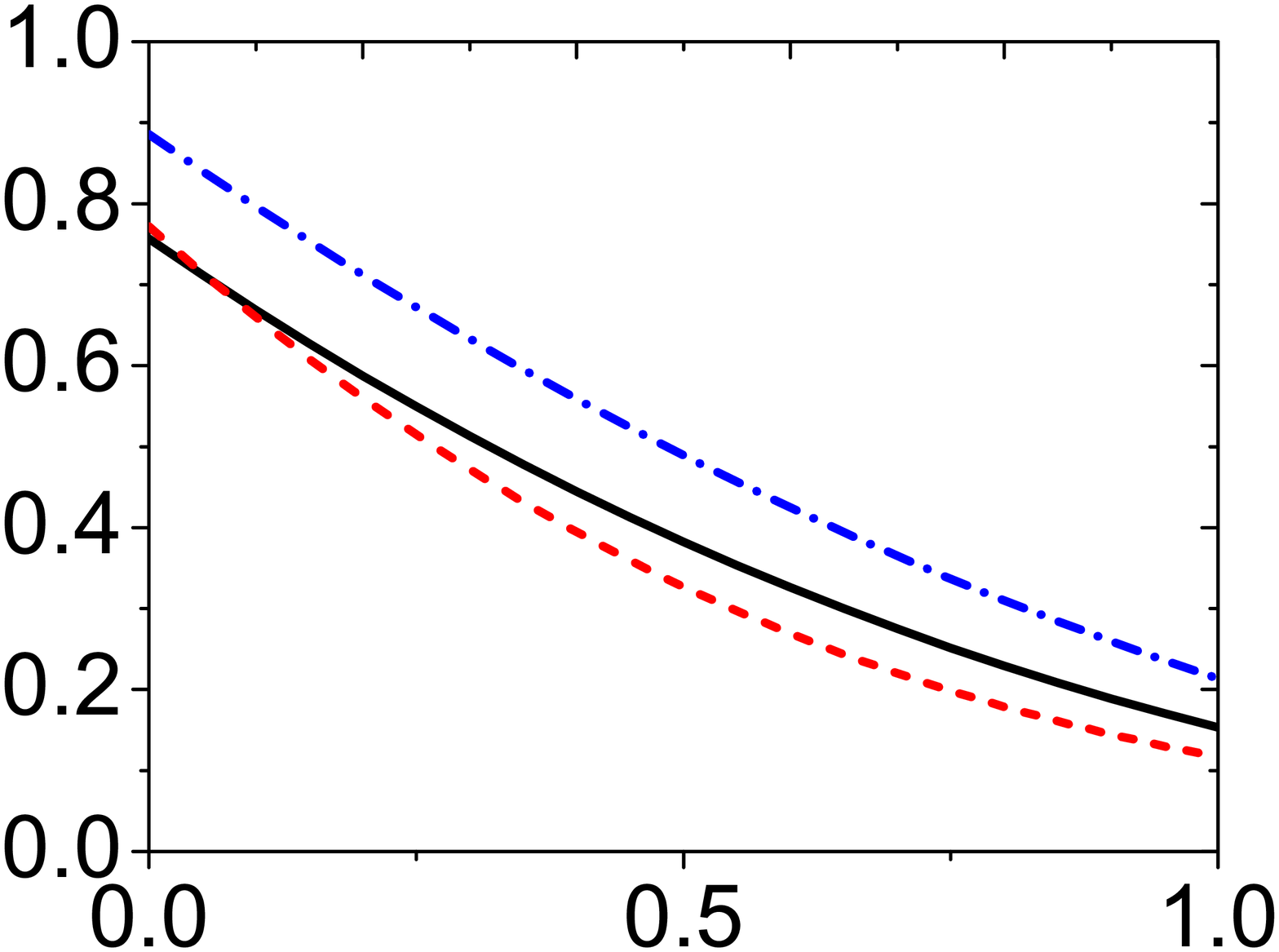}
  \put(-260,90){$\mathcal{F}(\rho_{g\ell}^{(f)})$}
 \put(-125,-1){\Large$\gamma$}
  \put(-60,145){ $(b)$}
     \caption{(a) The fidelity of the teleported state
     ${\mathcal{F}_{g\ell}}^{(1)}_{\phi^+}$(solid curves) and
     ${\mathcal{F}_{g\ell}}^{(0)}_{\phi^+}$ (dash-curves), where
     the channel strength, $p=0.0$. (b) The fidelity,
     ${\mathcal{F}_{g\ell}}^{(1)}_{\phi^+}$ for $p=0.1,0.3$ and
     $0.6$ for the solid, dash and dash-dot curves, respectively.}
       \end{center}
\end{figure}

\begin{table}
  \centering
  \begin{tabular}{|c|c|c|c|c|}
\hline
   Alice& Chirile & Bob &Fidelity\\
  \hline\hline
  $\rho_{\phi^+}$ & $1$ & $I$&${\mathcal{F}_{g\ell}}^{(1)}_{\phi^+}=\frac{1}{2}(\mu^4\mathcal{B}_1+\mu^2\nu^2(\mathcal{B}_4+
  \mathcal{B}_{13})+\nu^4\mathcal{B}_{16})$\\
 &0&$S_x$&${\mathcal{F}_{g\ell}}^{(0)}_{\phi^+}=\frac{1}{2}(\mu^4\mathcal{B}_6+\mu^2\nu^2(\mathcal{B}_7+
  \mathcal{B}_{10})+\nu^4\mathcal{B}_{11})$\\
  \hline
 $\rho_{\phi^-}$ & $1$ & $S_z$ & ${\mathcal{F}_{g\ell}}^{(1)}_{\phi^-}={\mathcal{F}_{g\ell}}^{(1)}_{\phi^+}$\\
 &0&$S_xS_z$&${\mathcal{F}_{g\ell}}^{(0)}_{\phi^-}={\mathcal{F}_{g\ell}}^{(0)}_{\phi^+}$\\
 \hline
  $\rho_{\psi^+}$ & $1$ & $S_x$& ${\mathcal{F}_{g\ell}}^{(1)}_{\psi^+}=
  \frac{1}{2}(\mu^4\mathcal{B}_{16}+\mu^2\nu^2(\mathcal{B}_4+
  \mathcal{B}_{13})+\nu^4\mathcal{B}_{1})$\\
  &0&$I$&${\mathcal{F}_{g\ell}}^{(0)}_{\psi^+}=\frac{1}{2}(\mu^4\mathcal{B}_{11}+\mu^2\nu^2(\mathcal{B}_7+
  \mathcal{B}_{10})+\nu^4\mathcal{B}_{6})$\\
  \hline
    $\rho_{\psi^-}$&$1$ & $S_xS_z$&${\mathcal{F}_{g\ell}}^{(1)}_{\psi^-}={\mathcal{F}_{g\ell}}^{(1)}_{\psi^+}$\\
  &0&$S_z$&${\mathcal{F}_{g\ell}}^{(0)}_{\psi^-}={\mathcal{F}_{g\ell}}^{(0)}_{\psi^+}$\\
  \hline
\end{tabular}
 \caption{ Teleportation protocol via decohered GHZ-like state as quantum channel}
\end{table}
The fidelity of the teleported state by using a decohered GHZ-like
state as quantum channel  is shown in Fig.(6). In Fig.(6a), we set
the channel strength $p=0$ and assume that Alice measures
$\rho_{\phi^+}$, while Charlie measures $"0"$ or $"1"$. The
general behavior  shows that the  fidelity  decays  as the damping
parameter $\gamma$ increases. However, the  decay rate of the
fidelity depends on the Charlie's measurements. For example, if
Charlie measures $"1"$, then the  rate decay  of the  fidelity
${\mathcal{F}_{g\ell}}^{(1)}_{\phi^+}$ is much smaller than that
depicted for   ${\mathcal{F}_{g\ell}}^{(0)}_{\phi^+}$, where
Charlie measures "0". Moreover, the fidelity
${\mathcal{F}_{g\ell}}^{(0)}_{\phi^+}$ vanishes completely at
$\gamma=1$, while ${\mathcal{F}_{g\ell}}^{(1)}_{\phi^+}$ doesn't.

Fig.(6b) shows the behavior of the fidelity
${\mathcal{F}_{g\ell}}^{(1)}_{\phi^+}$ for different values of the
channel strength $p$. In general, the fidelity decreases as
$\gamma$ increases. However,  for small values of of $p\in[0,5]$,
the fidelity decreases as $p$ increases. For $p>5$, the upper
values of the fidelity is larger than that depicted for small
values of $p$.

\section{Conclusion}
In this paper, we discussed the entanglement behavior of two
classes of tripartite entangled states, namely, GHZ and GHZ-like
states, passing through a generalized amplitude damping channel.
The effect of the channel strength and channel damping parameter
on the survival amount of entanglement are investigated. The decay
rate of entanglement  increases as the channel damping parameter
increases. However, the entanglement decreases faster for small
values of the channel strength, while for larger values the upper
bounds of entanglement are much larger.  It is shown that, the
behavior of entanglement of the GHZ-like state is more robust than
 GHZ state, where  the decay rate of entanglement for GHZ-like
state is smaller than that depicted for GHZ state.

The decohered  entangled tripartite  states  are used as quantum
channel to perform quantum teleportation. The fidelity of the
teleported state is investigated within and without the channel
strength. If the users use the decohered GHZ state as quantum
channel, the fidelity of the teleported state depends on the
channel's strength, channel's damping parameter and the analyzer's
angle.  It is shown that, the fidelity of the teleported state
decrease quickly as the analyzer angle increases. The decay rate
of the fidelity increases as the channel damping parameter
increases. Within larger values of the channel strength, the decay
rate increases and the fidelity decays faster. However, as the
channel damping parameter increases, the behavior of the fidelity
is  stable and fixed.

The possibility of using decohered GHZ-like state as quantum
teleportation is investigated. We show that the fidelity of the
teleportated state  depends on Bell measurements, Von Neumann, and
the channel's parameters. It is shown that, for some of the  Bell
measurements, the fidelity of the teleported state decays faster
and completely vanishes for larger values of the channel damping,
while decays  slowly for others and doesn't vanish as the channel
damping parameters increases. However, the decay rate of the
fidelity increases as the channel strength increases. This decay
can be decreased by increasing the channel's strength.

{\it In conclusion:}~Despit the generalized amplitude damping
channel causes a degradation of the entangled properties of the
initial entangled states, and consequently their efficiency to
perform teleportation. The upper bounds of entanglement and the
fidelity of the teleported states can be enhanced as one increases
the channel's strength. The phenomena of channel frozen doesn't
appear for tripartite state. However, for the fidelity of the
teleported state the frozen phenomena appears for larger values of
the channel strength. Finally, one can say that the  class of the
GHZ-like state is more robust than  GHZ states.

\end{document}